# Transverse instability at a compensated interface of ferromagnetic – antiferromagnetic bilayer

N. A. Usov

*Troitsk Institute for Innovation and Fusion Research, 142190, Troitsk, Moscow region, Russia*

Ching-Ray Chang and Zung-Hang Wei

*Department of Physics, National Taiwan University, Taipei, Taiwan, 10764, Republic of China*

The analytical theory for a ferromagnetic-antiferromagnetic bilayer with a compensated interface has been developed using an explicit expression for the interfacial interaction energy density. The stability of a solution describing domain walls within ferromagnetic and antiferromagnetic films has been investigated. It has been shown that even for high values of surface interaction strength a transverse instability originates within the antiferromagnetic film, because the deviation of the unit antiferromagnetic vector out of the interface plane becomes energetically favorable in large enough external magnetic field. However, it is possible to stabilize the spin distribution near the interface assuming additional in-plane anisotropy within the antiferromagnetic layer. In principle, this shows the way to control exchange bias just avoiding a complicated problem of domain rearrangement within the antiferromagnetic film.

## I. INTRODUCTION

The exchange coupling is observed for a variety of ferromagnetic (FM) and antiferromagnetic (AFM) materials being in atomic contact[1-3]. The effect is attributed to the exchange interaction between the FM and AFM spins at the interface. Recently, a number of important features of the AFM-FM exchange coupling has been successfully studied both experimentally and theoretically[2-10]. The most intriguing seems the so-called perpendicular exchange coupling existing at a compensated AFM surface, which is expected to have no net interfacial magnetic moment. This type of coupling was predicted by Koon[6] several years ago by means of numerical simulation. The existence of the perpendicular coupling at the compensated AFM-FM interface was also confirmed in numerical calculations of Schulthess and Butler[7,8], though they stated that this type of coupling did not necessarily lead to the exchange bias. These theoretical predictions are in agreement with the experiments on $FeF_2$/Fe and $MnF_2$/Fe bilayers[11-14] with (110) and (101) compensated orientations at the interface. The perpendicular coupling has also been discovered in $Fe_3O_4$/CoO bilayers with compensated (001) CoO surface[15-17].

Based on Heisenberg Hamiltonian for an AFM-FM bilayer it has been shown recently[18] that certain periodic deviations of the AFM and FM spins are energetically favorable near a compensated surface of AFM. The amplitude of the deviations decreases exponentially both into AFM and FM volumes. Therefore, macroscopically the AFM-FM exchange interaction can be considered as a surface magnetic anisotropy. In the absence of external magnetic field the surface interaction causes the FM spins to rotate perpendicular to the direction of the AFM spins at the interface. Under certain conditions the exchange coupling holds also in high enough external magnetic field making physics of the AFM-FM spin system so interesting.

In this paper we have developed an analytical theory for AFM-FM bilayer with a compensated interface using an explicit expression[18] for the energy density of the surface magnetic anisotropy. Due to the flat geometry of the problem a one-dimensional (1D) solution of the corresponding micromagnetic equations has to be studied. It describes the behavior of FM and AFM domain walls near the interface. We investigate also the stability of the 1D solution obtained. This investigation shows that the properties of the bilayer depend crucially on the strength of the surface interaction. If surface interaction strength is small, the exchange coupling at the interface breaks in moderate magnetic field. As a result the bilayer exhibits usual hysteretic behavior in external magnetic field. However, its coercive force increases with respect to that of free FM layer due to the influence of the interface interaction. On the other hand, for sufficiently high values of surface interaction strength formal 1D solution exists in arbitrary high external magnetic field. It describes purely reversible behavior of the bilayer. This is because the energy stored in AFM and FM domain walls returns back when external magnetic field decreases to zero. This regime would correspond to the exchange biasing if AFM and FM domain walls were stable against small perturbations. It is found however that transverse instability happens in AFM layer at certain critical magnetic field, when the deviation of the unit AFM vector out of the interface plane becomes energetically favorable.

It should be noted that in experiments on $FeF_2$/Fe and $MnF_2$/Fe bilayers[11-14], as well as for $Fe_3O_4$/CoO bilayer[15-17], the AFM film is certainly non-uniform. In the case of fluoride bilayers AFM film consists of twin domains with various directions of the easy anisotropy axes at the interface[13,19]. Twin domains exist also in CoO and other similar AFM oxides[20,21]. The condition for the onset of the transverse instability can hardly be satisfied for different domains simultaneously. In other words, in a real experiment the exchange coupling at the interface never breaks completely. However, the rearrangement of the AFM domains during the rotation of the FM magnetization in external magnetic field considerably complicates the experimental situation.

In this paper we have analyzed another possibility to avoid transverse instability at the interface, at least in principle. It has been shown that AFM spin distribution can be stabilized by means of additional in-plane anisotropy within the AFM layer. This possibility seems attractive as it enables one to control exchange bias just avoiding the complicated problem of the domain rearrangement in the AFM film.

This paper is organized as follows. In Section II the basic micromagnetic equations are stated and 1D solution for the AFM and FM domain walls is studied. The investigation of the stability of the 1D solution is presented in Section III. Section IV is devoted to the discussion of the results and conclusions.

## II. 1D SOLUTION

In this Section we study the influence of surface exchange interaction on the properties of a thin AFM-FM bilayer. Suppose that AFM and FM films correspond to the regions $0 \leq z \leq L_z$ and $L_z \leq z \leq L_z + d_z$ of the Cartesian coordinates, respectively. Due to a shape anisotropy a unit FM vector $\alpha$ is parallel to the film plane. Then the energy of the FM layer per unit square is given by

$$\frac{W_f}{S} = \int_{L_z}^{L_z+d_z} dz \left\{ \frac{C_f}{2}\left[\left(\frac{d\alpha_x}{dz}\right)^2 + \left(\frac{d\alpha_y}{dz}\right)^2\right] + K_f\left(1-(\alpha n)^2\right) - M_s \alpha H_0 \right\}. \quad (2.1)$$

Here $S$ is the square of the interface, $C_f$ and $K_f > 0$ are the exchange and the anisotropy constants of the FM layer, respectively, $M_s$ is the saturation magnetization, $H_0$ is the external magnetic field. The easy anisotropy axis of the FM layer is supposed to be parallel to the unit vector $n$.

Macroscopically, AFM layer is described by means of unit magnetization vectors of two sublattices, $\beta_1$ and $\beta_2$. A strong exchange interaction within the AFM causes the sublattices to be opposite, $\beta_2 = -\beta_1$. Then, the AFM unit vector $\beta = (\beta_1-\beta_2)/2 = \beta_1$. Note, that according to microscopic approach[18] a certain canting of the AFM sublattices exists very close to the interface. It is accompanied also by a small periodic orientation deviation of the unit FM vector. But these deviations decrease exponentially into AFM and FM volumes. Therefore, one can neglect them from macroscopic point of view accepted in the present paper.

It is instructive to study first a situation when the rotation of the unit AFM vector is restricted within the interface plane. With this restriction in mind, the energy of the AFM layer is given by

$$\frac{W_a}{S} = \int_0^{L_z} dz \left\{ \frac{C_a}{2}\left[\left(\frac{d\beta_x}{dz}\right)^2 + \left(\frac{d\beta_y}{dz}\right)^2\right] + K_a \beta_y^2 \right\}. \quad (2.2)$$

Here $C_a$ is the AFM exchange constant, $K_a > 0$ is the anisotropy constant of AFM layer. The AFM easy anisotropy axis is supposed to be parallel to the $x$-axis.

The interaction energy density at the AFM-FM interface with a compensated AFM surface is given by[18]

$$\frac{W_{int}}{S} = K_s(\alpha\beta)^2; \quad K_s = \frac{J_f S_f^2 + J_a S_a^2}{17.6 J_f J_a a^2} J_{int}^2. \quad (2.3)$$

Here $K_s$ is the surface anisotropy constant, $J_f > 0$ and $J_a > 0$ are the FM and AFM exchange integrals, respectively. $J_{int}$ is the exchange integral for the AFM-FM interaction, $S_f$ and $S_a$ are the lengths of the FM and AFM spins, correspondingly, $a$ is the lattice period. The expression (2.3) corresponds to the case of FeF$_2$/Fe or MnF$_2$/Fe bilayer. For a Fe$_3$O$_4$/CoO bilayer similar equation is stated[18] with the difference that the numerical coefficient 17.6 in Eq. (2.3) is corrected for 9.3.

It is convenient to introduce a polar representation for the vectors $\alpha$, $\beta$, $n$ and $H_0$ (see inset in Fig. 1 for the angle definitions)

$$\alpha = \{\cos\varphi, \sin\varphi, 0\}; \quad \beta = \{\cos\psi, \sin\psi, 0\};$$
$$n = \{\cos\varphi_k, \sin\varphi_k, 0\}; \quad H_0 = \{\cos\varphi_H, \sin\varphi_H, 0\}, \quad (2.4)$$

the external magnetic field being applied within the interface plane. Then the reduced total energy of the bilayer can be rewritten as follows

$$\frac{W}{2K_a S \delta} = \frac{1}{2}\int_0^L dz'\left(\left(\frac{d\psi}{dz'}\right)^2 + \sin^2\psi\right) + \int_L^{L+d} dz'\left[\frac{\kappa}{2}\left(\frac{d\varphi}{dz'}\right)^2 + \frac{\xi}{2}\sin^2(\varphi-\varphi_k) - h_e\cos(\varphi-\varphi_H)\right] + \frac{\eta}{2}\cos^2(\psi_L - \varphi_L). \quad (2.5)$$

Here we introduce the dimensionless parameters

$$\kappa = \frac{C_f}{C_a}; \quad \xi = \frac{K_f}{K_a}; \quad h_e = \frac{M_s H_0}{2K_a}; \quad \eta = \frac{K_s}{K_a \delta}, \quad (2.6)$$

where $\delta = \sqrt{C_a/2K_a}$ has the meaning of the AFM domain wall width, $\psi_L = \psi(L)$, $\varphi_L = \varphi(L)$ denote the magnitudes of these angles at the interface $z' = L$. The reduced lengths introduced in Eq. (2.5) are given by $z' = z/\delta$; $L = L_z/\delta$; $d = d_z/\delta$.

Making a variation of the total energy (2.5) with respect to the angles $\psi$ and $\varphi$ one obtains both the differential equations

$$\frac{d^2\psi}{dz'^2} = \sin\psi\cos\psi; \quad (2.7a)$$

for $0 < z' < L$, and

$$\kappa \frac{d^2\varphi}{dz'^2} = \xi \sin(\varphi - \varphi_k)\cos(\varphi - \varphi_k) + h_e \sin(\varphi - \varphi_H),$$
(2.7b)

for $L < z' < L+d$, as well as the boundary conditions $d\psi/dz' = 0$ at $z' = 0$, $d\varphi/dz' = 0$ at $z' = L+d$, and

$$\frac{d\psi}{dz'} = \frac{\eta}{2}\sin 2(\psi_L - \varphi_L);$$
(2.8a)

$$\kappa \frac{d\varphi}{dz'} = \frac{\eta}{2}\sin 2(\psi_L - \varphi_L),$$
(2.8b)

at the interface $z' = L$.

The first integral of Eq. (2.7a) is given by

$$(d\psi/dz')^2 = \sin^2\psi + C.$$

The integration constant $C$ has to be equaled to zero to get a relation

$$d\psi/dz' = \sin\psi.$$
(2.9)

The latter satisfies approximately the boundary condition at $z' = 0$, because $\psi(z')$ decreases exponentially into the AFM volume, $z' < L$. Then the solution of Eq. (2.7a) is given by

$$\cos\psi = \tanh(z_0 - z');$$
$$z_0 = L - \frac{1}{2}\ln\frac{1-\cos\psi_L}{1+\cos\psi_L},$$
(2.10)

where $z_0$ is another constant of integration. It is determined by the value of $\psi_L$ at $z' = L$.

Using the relation (2.9) the boundary condition (2.8a) becomes

$$\sin\psi_L = \frac{\eta}{2}\sin 2(\psi_L - \varphi_L).$$
(2.11)

It is easy to get also an implicit solution of Eq. (2.7b), but it is more convenient to integrate the set of equations (2.7), (2.8) numerically using, for example, the algorithm developed for calculation of 1D domain wall in amorphous ferromagnetic wire[22].

Let us prove now that the behavior of the solutions of the set of equations (2.7), (2.8) depends crucially on the strength of the AFM-FM interaction $\eta$. Consider for simplicity a situation when $\varphi_k = \pi/2$, so that the FM easy axis is parallel to the y-axis. Then, in the absence of external magnetic field, $h_e = 0$, there are two stable solutions of Eqs. (2.7), (2.8), namely $\psi = 0$; $\varphi = \pi/2$ and $\psi = 0$; $\varphi = 3\pi/2$. (We put aside two other equivalent solutions with $\psi = \pi$). This is because the interaction energy (2.3) is minimal when the unit FM and AFM vectors are perpendicular to each other. Therefore, this situation corresponds to the so-called perpendicular coupling discovered recently in experiments with bilayers FeF$_2$/Fe and MnF$_2$/Fe[11-14], as well as Fe$_3$O$_4$/CoO[15-17] having compensated AFM-FM interface.

If external magnetic field increases from zero in a typical situation shown in the inset of Fig. 1, the unit FM vector will start rotating to the magnetic field direction and domain walls originate both in the FM and AFM layers, as shown in Fig. 1. This means that the angle $\psi_L$ increases as function of the external magnetic field. It follows from Eq. (2.11), that if $\eta < 2$ there is an ultimate angle, $\psi_{Lc} = \arcsin(\eta/2)$, for the rotation of the unit AFM vector at the interface. The corresponding critical magnetic field can be found from the relation $\psi_L(h_c) = \psi_{Lc}$. Strictly speaking, $h_c$ is an upper estimate for the actual coercive force of the bilayer, because at $h \geq h_c$ the initial stable solution certainly disappears. As a result, for $\eta < 2$ a bilayer shows usual hysteretic behavior. It is demonstrated in Fig. 2a for $\eta = 0.58$ and $\eta = 1.15$, respectively. One can see from Fig. 2a that in the case of $\eta < 2$ the influence of interaction (2.3) only leads to the increase of the bilayer coercive force.

On the contrary, if $\eta \geq 2$ the initial solution exists even in very high external magnetic field. In this case the behavior of the bilayer is purely reversible, as shown in Fig. 2b, where similar calculation is carried out for $\eta = 2.31$. The physical reason for the reversible behavior is evident. If magnetic field increases, the energy of the bilayer will be stored within the FM and AFM domain walls. This energy returns back when magnetic field decreases. This process is purely reversible if there are no imperfections in the AFM and FM layers. Therefore, for perfect layers the shifted 'hysteresis' loop has zero square. This conclusion resembles the well-known result of Koon[6], who discovered similar behavior of a bilayer with a compensated interface by means of numerical simulation under the condition that the spin rotations are restricted within the interface plane.

The non-hysteretic magnetization curve in Fig. 2b turns out to be shifted from the origin by the 'exchange bias' field, $h_{eb}(\varphi_H)$. The latter depends on the external magnetic field direction, $\varphi_H > \pi$. It follows from Eqs. (2.7), (2.8) and (2.11) that for a thin FM layer, $d \leq 1$, in the limit of $\xi << 1$ the highest exchange bias field for the present model is given by $h_{eb}(3\pi/2) = (1-1/\eta^2)^{1/2}/d$.

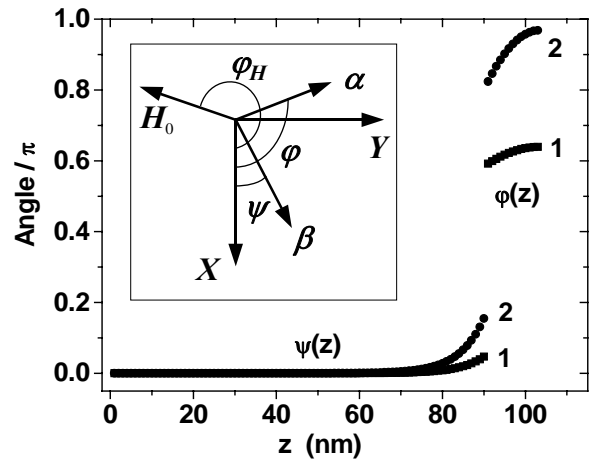

FIG. 1. The domain walls in AFM, $\psi(z)$, and FM, $\varphi(z)$, at different values of external magnetic field: 1) $H_0 = 400$ Oe; 2) $H_0 = 800$ Oe. The magnetic field direction $\varphi_H = 1.4\pi$, interaction strength $\eta = 1.15$. Inset shows the arrangement of the unit AFM and FM vectors and definition of the angles $\psi$, $\varphi$, and $\varphi_H$.

The 1D calculations presented in Figs. 1, 2 have mostly illustrative meaning, because in real experiments[11-17] the AFM layer is certainly non-uniform. Rather, it consists of tiny domains with different directions of the easy anisotropy axes at the interface plane. Evidently, this fact complicates the experimental situation. Nevertheless, it seems that the results presented in this Section shed light on a possible origin of the exchange bias in a bilayer with a compensated interface. For the calculations presented in Figs. 1, 2 we use values typical for the FeF$_2$/Fe bilayer. Namely, the anisotropy constants $K_a = 3 \cdot 10^6$ erg/cm$^3$ and $K_f = 10^3$ erg/cm$^3$ have been used for FeF$_2$ and for soft Fe film, respectively[12]. The exchange constants are assumed to be $C_f = 2.3 \cdot 10^{-6}$ erg/cm for Fe and $C_a = 2 \cdot 10^{-6}$ erg/cm for FeF$_2$. For FeF$_2$/Fe bilayer surface anisotropy constant has been estimated[18] to be of the order of $K_s \sim 1$ erg/cm$^2$. To study the influence of the surface anisotropy constant on the bilayer properties the calculations in Figs 2a, 2b have been made for the cases $K_s = 1$, 2 and 4 erg/cm$^2$, respectively. It corresponds to the values of $\eta = 0.58$, $\eta = 1.15$ and $\eta = 2.31$ as indicated in the corresponding figures. The thickness of the AFM and FM layers are given by $L_z = 90$ nm and $d_z = 13$ nm[12] for all of the cases investigated.

The reduced magnetic anisotropy of the FM layer turns out to be very small, $\xi \approx 3 \cdot 10^{-4}$, for the data presented in Figs. 1, 2. It is interesting to note, that the behavior of the bilayer becomes *partly* irreversible, if one assumes sufficiently high value of the magnetic anisotropy in the FM layer, $\xi > \xi_c \sim 1$. It can be shown that in this case additional stable solution of Eqs. (2.7), (2.8) appears in high enough external magnetic field. One can see in Fig. 3 that the shifted hysteresis loops of the bilayer in the case of $\xi = 1$ have finite square, with a finite value of the coercive force. Though so large values of parameter $\xi$ are not reliable in the experiment, this example shows that additional FM anisotropy may lead to a shifted hysteresis loop with a finite square.

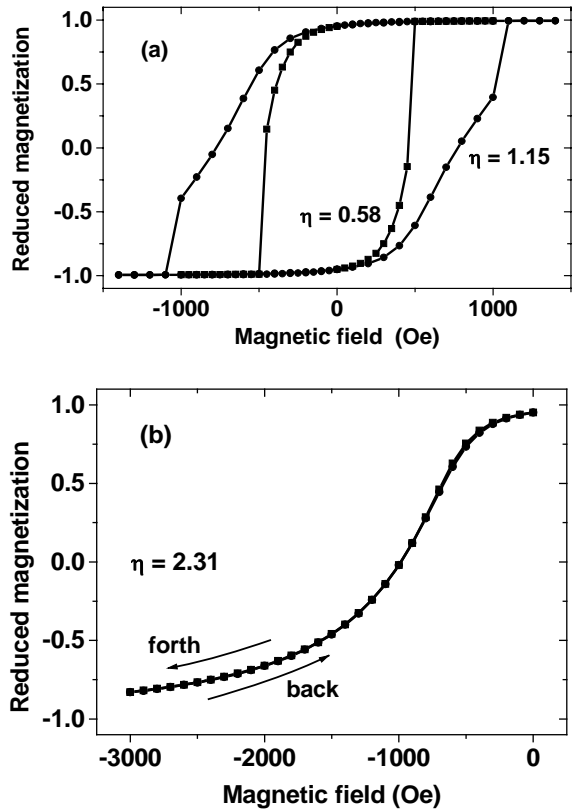

FIG. 2. The behavior of AFM-FM bilayer with a compensated interface at different exchange interaction strengths: a) hysteresis loops for $\eta < 2$; b) non hysteretic magnetization curve for $\eta \geq 2$.

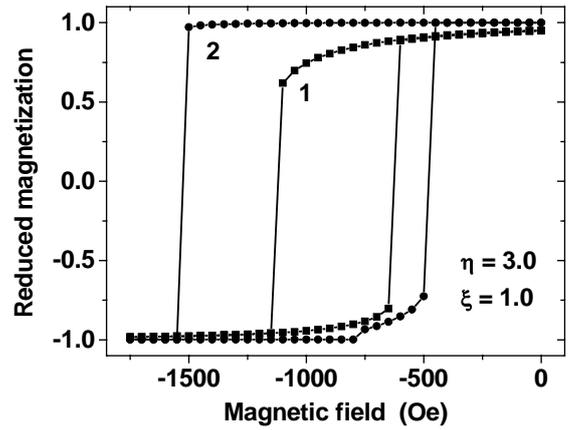

FIG. 3. Shifted hysteresis loops for a bilayer with a sufficiently high value of magnetic anisotropy in the FM film at different direction of external magnetic field: 1) $\varphi_H = 1.4\pi$; 2) $\varphi_H = 1.49\pi$.

The 1D calculations presented in this Section demonstrate possible physical reason for the exchange biasing in a bilayer with a compensated AFM-FM interface. They show also that exchange interaction at the interface has to be sufficiently strong, $\eta \geq 2$, to be able to maintain exchange coupling of the AFM and FM layers in high enough external magnetic field. However, to consider 1D calculations seriously one has to investigate their stability.

### III. TRANSVERSE INSTABILITY

To investigate a stability of the 1D solution let us simplify initial functional (2.5) to be able to carry out explicit calculation. One can take into account that in many experimental situations FM layer thickness is small and comparable with the FM exchange length. In this case one can neglect $z$-dependence of the unit FM vector setting in Eq. (2.5) $\varphi(z) \approx \varphi$. For a soft FM layer one can neglect also magnetic anisotropy in this layer. On the other hand, shape anisotropy of the FM layer is high due to large value of the saturation magnetization. Therefore, the unit FM vector is still restricted to rotate within the interface plane. On the other hand, in this Section we will not impose any constraint on the direction of the unit AFM vector. Therefore, let us study the following functional

$$\frac{W}{S} = \int_0^{L_z} dz \left[ \frac{C_a}{2} \left( \frac{d\boldsymbol{\beta}}{dz} \right)^2 + K_a \left( \beta_y^2 + \beta_z^2 \right) \right] +$$

$$K_s(\boldsymbol{\alpha}\boldsymbol{\beta})^2 - M_s d_z \boldsymbol{\alpha} \boldsymbol{H}_0. \quad (3.1)$$

Using the spherical representation for the unit AFM vector

$$\boldsymbol{\beta} = \{\sin\omega\cos\psi, \sin\omega\sin\psi, \cos\omega\}, \quad (3.2)$$

and Eqs. (2.4) for the vectors $\boldsymbol{\alpha}$ and $\boldsymbol{H}_0$, the reduced total energy of the bilayer (3.1) is given by

$$\frac{W}{2K_a S\delta} = \frac{1}{2}\int_0^L dz'\left[\sin^2\omega\left(\left(\frac{d\psi}{dz'}\right)^2 + \sin^2\psi\right) + \left(\frac{d\omega}{dz'}\right)^2 + \cos^2\omega\right] +$$
$$\frac{\eta}{2}\sin^2\omega_L \cos^2(\psi_L - \varphi) - h_e d \cos(\varphi - \varphi_H), \quad (3.3)$$

where $\omega_L = \omega(L)$, $\psi_L = \psi(L)$ are the magnitudes of these angles at the interface.

If rotation of the unit AFM vector is restricted within the interface plane, then $\omega(z') = \pi/2$. To study a possibility of deviation of vector $\boldsymbol{\beta}$ out of the interface plane one can make in Eq. (3.3) a substitution $\omega(z') \approx \pi/2 + \varepsilon(z')$. Here a perturbation $\varepsilon(z')$ is supposed to be small, $|\varepsilon(z')| \ll 1$. In the lowest order to the perturbation $\varepsilon$ the total energy (3.3) can be represented as a sum of the contributions, $w = w^{(0)} + w^{(1)}$, where

$$w^{(0)} = \frac{1}{2}\int_0^L dz'\left[\left(\frac{d\psi}{dz'}\right)^2 + \sin^2\psi\right] +$$
$$\frac{\eta}{2}\cos^2(\psi_L - \varphi) - h_e d\cos(\varphi - \varphi_H), \quad (3.4)$$

and

$$w^{(1)} = \frac{1}{2}\int_0^L dz'\left[\left(\frac{d\varepsilon}{dz'}\right)^2 + \left(\cos^2\psi - \left(\frac{d\psi}{dz'}\right)^2\right)\varepsilon^2\right] -$$
$$\frac{\eta}{2}\varepsilon_L^2 \cos^2(\psi_L - \varphi), \quad (3.5)$$

respectively. In Eq. (3.5) we denote $\varepsilon_L = \varepsilon(L)$. Making a variation of the functional (3.4) with respect to the angles $\psi$ and $\varphi$ one arrives to the same Eq. (2.7a) within the interval $0 < z' < L$, with the boundary conditions $d\psi/dz' = 0$ at $z' = 0$ and

$$\sin\psi_L = \frac{\eta}{2}\sin 2(\psi_L - \varphi); \quad (3.6a)$$

$$h_e d \sin(\varphi_H - \varphi) = \frac{\eta}{2}\sin 2(\psi_L - \varphi), \quad (3.6b)$$

at the interface $z' = L$. Note, that in Eq. (3.6a) the relation (2.9) has been used. The same solution (2.10) for the AFM domain wall holds here. Therefore, at every given value of external magnetic field $h_e$ the equilibrium values of the angles $\psi_L$ and $\varphi$ can be determined from the set of Eqs. (3.6). It is worth noting that the rotation of the unit AFM vector with large values of $\psi_L$ at the interface is again possible under the condition $\eta \geq 2$.

To check a stability of this solution one has to analyze the functional (3.5). Using in Eq. (3.5) the relation (2.9) one obtains

$$w^{(1)} = \frac{1}{2}\int_0^L dz'\left[\left(\frac{d\varepsilon}{dz'}\right)^2 + \cos(2\psi)\varepsilon^2\right] - \frac{\eta}{2}\varepsilon_L^2 \cos^2(\psi_L - \varphi). \quad (3.7)$$

Let us minimize the integral term in this equation under a condition that function $\varepsilon(z')$ has a prescribed value $\varepsilon_L$ at the interface. This means that the deviation $\delta\varepsilon(z') = 0$ at $z' = L$. Under this condition the variation of the integral term in Eq. (3.7) leads to the differential equation

$$\frac{d^2\varepsilon}{dz'^2} = \cos(2\psi)\varepsilon; \qquad 0 < z' < L, \quad (3.8)$$

with the only boundary condition $d\varepsilon/dz' = 0$ at $z' = 0$. It can be checked by substitution that Eq. (3.8) has the first integral $d\varepsilon/dz' = \cos(\psi)\varepsilon$. The second integral of this equation is given by $\varepsilon = C\sin\psi$, where $C$ is an arbitrary constant. In view of the condition $\psi \to 0$ at $z' = 0$, we have also $\varepsilon \to 0$ at $z' = 0$, so that the boundary condition to the Eq. (3.8) is satisfied. Using Eq. (3.8) in Eq. (3.7) one obtains

$$w^{(1)} = \frac{1}{2}\int_0^L dz' \frac{d}{dz'}\left[\varepsilon\left(\frac{d\varepsilon}{dz'}\right)\right] - \frac{\eta}{2}\varepsilon_L^2 \cos^2(\psi_L - \varphi) =$$
$$\frac{1}{2}\left[\cos\psi_L - \eta\cos^2(\psi_L - \varphi)\right]\varepsilon_L^2.$$

This expression can be further simplified using Eq. (3.6a). In this way one gets finally

$$w^{(1)} = \frac{1}{2}f(\psi_L)\varepsilon_L^2;$$
$$f(\psi_L) = \cos\psi_L - \frac{\eta}{2} + \sqrt{\left(\frac{\eta}{2}\right)^2 - (\sin\psi_L)^2}. \quad (3.9)$$

It follows from Eqs. (3.6) that the angle $\psi_L$ is small at small enough values of $h_e$. Then the coefficient $f(\psi_L) > 0$, so that the functional $w^{(1)}$ is definitively positive. However, the angle $\psi_L$ increases as function of $h_e$. It is easy to see that $f(\psi_L)$ becomes negative if $\psi_L$ exceeds a critical angle

$$\sin\psi_{tr} = \sqrt{1 - 1/\eta^2}. \quad (3.10)$$

This means that the perturbation $\varepsilon(z') \sim \sin\psi$ is energetically favorable at $\psi_L \geq \psi_{tr}$. The condition $\psi_L = \psi_{tr}$ is the onset of the transverse instability at the surface of the AFM layer. The critical angle is minimal, $\psi_{tr} = 60°$, at $\eta = 2$, but it approaches to 90° for $\eta \gg 1$. Due to the transverse instability the unit AFM vector deviates from the interface plane, so that the initial exchange coupling at the interface breaks.

It is worth noting that Eq. (3.9) is valid also for $\eta < 2$, within the interval $2^{1/2} < \eta < 2$. Therefore, both nucleation modes, transverse instability and the uniform rotation of vector $\boldsymbol{\beta}$, have to be taken into account in this interval of $\eta$.

The first term in the function $f(\psi_L)$ describes the stability of the AFM domain wall itself. It is positive at $\psi_L < \pi/2$. However, it becomes negative for $\psi_L > \pi/2$ showing that the AFM domain wall is unstable in this interval. This instability is similar to that one discussed by W.F. Brown, Jr. many years ago[23]. The second two terms in the expression for $f(\psi_L)$ originate due to AFM-FM interaction (2.3). The sum of these terms is negative for all values of $\psi_L$ within the interval $0 < \psi_L < \pi$. Therefore, the AFM-FM interaction only decreases the threshold for the transverse instability. This is because the interaction energy (2.3) decreases due to deviation of vector $\boldsymbol{\beta}$ out of the interface plane.

It follows from the above discussion that the 1D results presented in Section II may have some meaning only if the transverse instability at the AFM-FM interface can be somehow avoided. As we mentioned above, in real experiment the AFM layer consists of tiny twin domains. The directions of the easy anisotropy axes in different domains are perpendicular to each other at the interface[12,13]. Therefore, even if the condition for the onset of the transverse instability is fulfilled for the one type of domains, it cannot be satisfied for the other type of domains having perpendicular direction of the easy anisotropy axis. One can assume that this leads to a stabilization of the spin distribution in the AFM layer near the interface. The assumption has been confirmed recently[24] by means of numerical simulation of the magnetization process in AFM-FM bilayer having randomly distributed AFM domains within the AFM film. This shows an important role of the AFM domains in the exchange biasing. Actually, it is well recognized now[19,25] that it is the rearrangement of the AFM domains during the rotation of the FM magnetization in the external magnetic field that makes the experimental situation for $FeF_2/Fe$ and $MnF_2/Fe$ bilayers so complicated.

It is interesting to consider another possibility to avoid transverse instability at the interface of AFM-FM bilayer. As we have seen above, even in case of $\eta \gg 1$ the transverse instability happens due to the AFM domain wall instability at angles $\psi_L > \pi/2$. However, the out-of-plane rotation of the vector $\boldsymbol{\beta}$ may become energetically unfavorable if one assumes additional in-plane anisotropy within the AFM layer. Let us modify slightly Eq. (3.1) introducing additional in-plane anisotropy for the AFM film with the energy density $K_{a1}\beta_z^2$. Evidently, this energy contribution will stabilize in-plane AFM spin distribution if anisotropy constant $K_{a1}$ is positive and high enough. Using the dimensionless parameter $k = K_{a1}/K_a$ one can easily check that Eq. (3.4) remains unchanged, whereas Eq. (3.5) becomes

$$w^{(1)} = \frac{1}{2}\int_0^L dz' \left[\left(\frac{d\varepsilon}{dz'}\right)^2 + \cos(2\psi)\varepsilon^2 + k\varepsilon^2\right] - \frac{\eta}{2}\varepsilon_L^2 \cos^2(\psi_L - \varphi). \quad (3.11)$$

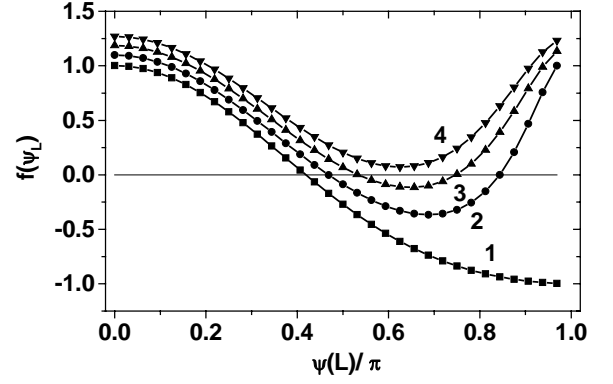

FIG. 4. The function $f(\psi_L)$ at different values of the parameter $k = K_{a1}/K_a$: 1) $k = 0$; 2) $k = 0.2$; 3) $k = 0.4$; 4) $k = 0.6$.

The corresponding variational problem can hardly be solved analytically. Nevertheless, a simple estimation

$$\int_0^L dz' \left[\left(\frac{d\varepsilon}{dz'}\right)^2 + \cos(2\psi)\varepsilon^2 + k\varepsilon^2\right] \geq$$
$$\int_0^L dz' \left[\left(\frac{d\varepsilon}{dz'}\right)^2 + (k-1)\varepsilon^2\right] \geq \sqrt{(k-1)}\varepsilon_L^2$$

shows that the integral term in Eq. (3.11) becomes definitely positive for $k > 1$. Then, in view of Eq. (3.9), a stability criterion for the AFM-FM spin distribution near the interface is given by

$$k > k_c = \frac{\eta^2}{2}\left(1 - \sqrt{1 - (2/\eta)^2}\right). \quad (3.12)$$

In fact, the estimation (3.12) is too coarse. Based on Eq. (3.11) for every given value of the interaction strength $\eta$ one can find more accurate estimation for $k_c$ numerically by means of the factorization method[26]. For example, Fig. 4 shows the behavior of the function $f(\psi_L)$ for different values of the parameter $k$ at $\eta = 4$. Note, that curve 1 in Fig. 4 coincides with Eq. (3.9). It can be seen that the condition $f(\psi_L) > 0$ is certainly fulfilled for $k \geq 0.6$. This value is about 40% lower then that one given by Eq. (3.12), $k_c = 1.07$.

TABLE 1. Reduced critical strength of the in-plane anisotropy

| $\eta$ | 2.5 | 3.0 | 3.5 | 4.0 | 4.5 | 5.0 |
|---|---|---|---|---|---|---|
| $k_{c,\,num}$ | 0.77 | 0.63 | 0.57 | 0.52 | 0.49 | 0.47 |

Table 1 summarizes the results of numerical investigation of the stability criterion for the functional (3.11) at different values of the bilayer interaction strength. One can see that moderate values of the anisotropy ratio $k$ are sufficient to stabilize AFM domain wall and to preserve exchange coupling at the interface within the whole interval of the angles $0 \leq \psi_L \leq \pi$.

## IV. DISCUSSION AND CONCLUSIONS

It has been shown recently[18] that exchange interaction at the interface of a bilayer with a compensated AFM surface is equivalent to a surface magnetic anisotropy. The surface interaction energy is minimal when the unit FM and AFM vectors are perpendicular to each other at the interface. This leads to a perpendicular exchange coupling in accordance with the experimental findings for the bilayers $FeF_2$/Fe, $MnF_2$/Fe[11-14], as well as for $Fe_3O_4$/CoO bilayers with compensated (001) CoO interface[15-17]. When FM spins rotate under the influence of external magnetic field the interface exchange coupling causes the AFM spins to follow this rotation. Thus, the FM and AFM domain walls originate close to the interface. It has been shown that the evolution of the AFM-FM spin system in external magnetic field depends crucially on the AFM-FM interaction strength $\eta$. If $\eta < 2$ the rotation of the unit AFM vector at the interface is restricted by a critical angle, $\psi_{Lc} = \arcsin(\eta/2)$, so that the exchange coupling breaks at $\psi(L) \geq \psi_{Lc}$. As a result, the bilayer shows usual hysteretic behavior, though the AFM-FM interaction increases its coercive force. On the other hand, if $\eta \geq 2$ the unit AFM vector formally can rotate up to large angles $\psi(L) \approx \pi$ at the interface. This rotation is reversible, because the energy stored within the FM and AFM domain walls returns back when external magnetic field decreases to zero. The reversible rotation of the FM and AFM spins would lead to the exchange biasing if the AFM-FM spin system were stable during evolution in external magnetic field up to high enough values of $h_e$. However, it has been proved analytically that for AFM layer with uniaxial magnetic anisotropy AFM domain wall looses its stability at $\psi(L) > \pi/2$. This is because the deviation of the unit AFM vector out of the interface plane turns out to be energetically favorable for large rotation angles at the interface.

Note, the AFM layer is certainly non-uniform in the experiment. It consists of tiny twin domains with perpendicular directions of easy anisotropy axes for different type of domains[12,13]. Therefore, the conditions for the onset of the transverse instability can hardly be satisfied for both domain groups simultaneously. In this case the exchange coupling at the AFM-FM interface never breaks completely. This conclusion has been confirmed recently[24] by means of numerical simulation of the magnetization process in AFM-FM bilayer with randomly distributed domains in the AFM film.

In this paper we have considered another possibility to avoid transverse instability in the AFM layer, at least in principle. It is shown that AFM domain wall can be stabilized by means of additional in-plane magnetic anisotropy within the AFM layer. For this case the stability criterion is obtained numerically as function of the interaction strength $\eta$. It is found that moderate values of the AFM anisotropy ratio, $k \sim 0.5 - 0.6$, are sufficient to stabilize AFM domain wall near the interface and to maintain the exchange coupling in arbitrary high applied magnetic field. One may hope that it will be easier to control exchange bias in a bilayer with a uniform AFM film just avoiding all problems with the rearrangement of the AFM domains. Therefore, it would be interesting to check this possibility in the experiment.

---